\documentclass[journal]{IEEEtran}

\usepackage{cite}
\usepackage{graphicx}
\usepackage{amsmath}
\usepackage{array}
\usepackage{booktabs}
\usepackage{multirow}
\usepackage{url}

\usepackage{tabularx}

\newcounter{listing}
\renewcommand{\thelisting}{\arabic{listing}}

\newcommand{\asmcode}[1]{{\ttfamily\footnotesize #1}}
\newcommand{\asmindent}{\hspace*{1.7em}}

\begin{document}

\title{SEAM-V: A Hybrid-Decoupled RISC-V Vector Processor with Backend-Visible Packet Semantics and Source-Lifetime-Aware Scheduling}

\author{Weiying~Wang
        and~Zhiwei~Zhang
\thanks{Weiying Wang and Zhiwei Zhang are with the Institute of Automation, Chinese Academy of Sciences, Beijing 100049, China, and also with the School of Artificial Intelligence, University of Chinese Academy of Sciences, Beijing 100049, China.}
\thanks{Corresponding authors: Weiying Wang (e-mail: wangweiying2021@ia.ac.cn).}
}

\maketitle

\begin{abstract}
Data-parallel workloads in deep learning and scientific computing continue to increase the demands on processor throughput, energy efficiency, and scalability. The RISC-V Vector Extension (RVV) supports scalable execution through a vector-length-agnostic model, yet many tightly coupled implementations still rely on the scalar core to supply vector instructions individually and are therefore constrained by instruction supply, scalar-side progress, memory
stalls, and conservative dependence management in short-vector, loop-tail, and control/memory-interleaved scenarios. This paper presents SEAM-V, a hybrid-decoupled RVV processor that uses task-level decoupling, local instruction supply, and VLIW-style packing to form a continuous stream of execute packets (EPs). During EP formation and request serialization, SEAM-V preserves the association between prefetch intent and the corresponding load to support request-bound prefetching, while lane-level source-read completion is used to release pure write-after-read (WAR) dependences early; other dependences remain governed by conventional mechanisms. Relative to an Ara-based tightly coupled baseline (TC), SEAM-V achieves a geometric mean speedup of \(1.38\times\) across 17 representative kernel configurations. The one-dimensional vector, BLAS and matrix, and fixed-size application workload groups achieve \(1.56\times\), \(1.35\times\), and \(1.23\times\), respectively. Synthesis and power analysis show that SEAM-V increases total cell area by 4.29\% and geometric mean runtime power by 17.30\%, while reducing task energy by 12.70\%, demonstrating improved sustained execution efficiency and task-level energy efficiency with limited area overhead.
\end{abstract}

\IEEEpeerreviewmaketitle

\section{Introduction}
\label{sec:introduction}

\IEEEPARstart{M}{any} workloads in deep neural networks, scientific computing, and autonomous driving exhibit substantial data-level parallelism and contain phases of regular computation or streaming memory access~\cite{fathom_reference_workloads_deep_learning,berkeley_parallel_computing_landscape}. Their throughput demands continue to drive general-purpose processors toward vector architectures~\cite{araxl_physically_scalable_ultra_wide_risc_v_vector_processor}. The RISC-V Vector Extension (RVV) provides an open, scalable, and vector-length-agnostic programming model, offering a unified interface for software portability and performance scaling across implementations with different physical vector lengths~\cite{risc_v_vector_extension_1_0}. Many existing RVV processors adopt a tightly coupled organization in which the scalar core handles instruction fetch and decode, control flow, address and loop-state maintenance, and vector instruction issue, while the vector unit executes data-parallel commands~\cite{ara_1ghz_scalable_energy_efficient_risc_v_vector_processor,ara2_exploring_single_multi_core_vector_processing,spatz_clustering_compact_risc_v_vector_units,risc_v2_scalable_vector_processor}. Although this organization provides a clean interface and broad compatibility, vector execution is driven by the scalar core one instruction at a time, making it difficult to sustain a continuous stream of vector requests at a sufficient rate. Completion-based dependence management may also retain source register read
claims longer than necessary, limiting backend scheduling opportunities.

These limitations appear across different vector granularities and kernel types. Short vectors, loop tails, small-batch operators, and small-matrix computations cannot effectively amortize the overheads of scalar control, address updates, and instruction issue, thereby exposing instruction throughput bottlenecks. Kernels with frequent control and address updates are constrained by scalar/vector interleaving, while regular but memory-sensitive kernels may experience execution gaps because of memory stalls and conservative dependence handling. Even when an application exhibits substantial data-level parallelism, its efficiency remains jointly affected by vector granularity, control intensity, memory behavior, and dependence management. Thus, bottlenecks in tightly coupled RVV processors are not confined to short-vector scenarios; across different kernels, they manifest as insufficient instruction supply, constrained scalar-side progress, and reduced sustained execution efficiency under memory and dependence constraints~\cite{troop_at_the_roofline_performance_vector_processors}.

Existing approaches improve vector execution efficiency by strengthening instruction supply and dynamic scheduling~\cite{out_of_order_vector_architectures,instruction_scheduling_saturn_vector_unit}, optimizing vector backends and memory paths~\cite{troop_at_the_roofline_performance_vector_processors}, or adopting decoupled or streaming execution~\cite{implementing_scale_vector_thread_processor,hwacha_vector_fetch_architecture_manual,decoupled_access_execute_computer_architectures,stream_semantic_registers,snitch_tiny_pseudo_dual_issue_processor}. VLIW and EPIC architectures instead use software- or compiler-generated execution packets to explicitly organize local instruction-level parallelism~\cite{instruction_level_parallel_processing_history_overview_perspective,epic_explicitly_parallel_instruction_computing}. Our prior work further proposed a VLIW-driven hybrid-decoupled vector architecture in which, after the Host submits a task, the vector coprocessor autonomously advances an interleaved scalar/vector instruction stream through local instruction supply, scalar execution, and HINT-guided execute-packet formation~\cite{boosting_vector_instruction_throughput_risc_v_hybrid_decoupled_vliw}. However, in that architecture, prefetch intent carried by a HINT may still be lost when an EP is serialized into individual vector requests. Conventional vector dependence management also typically retains source register read claims until the reading instruction completes, without exploiting runtime state indicating that all required source data may have been captured earlier. Therefore, the core problem addressed in this paper is how to preserve packet-derived memory intent across packet-to-request serialization and use actual source operand lifetimes to reduce unnecessary dependence waits while retaining the runtime execution capability of the vector backend.

Building on this hybrid-decoupled architecture, this paper presents SEAM-V, which uses local instruction supply and VLIW-style packing to organize an interleaved scalar/vector instruction stream into a continuous stream of execute packets (EPs), with each EP serving as the basic unit of local instruction organization and hybrid dispatch. Software uses lightweight HINT headers to provide packing permissions, local loop markers, and memory access intent. SEAM-V preserves the association between prefetch intent and the corresponding load throughout EP formation, hybrid dispatch, and request serialization and generates request-bound prefetch context, thereby realizing backend-visible packet-derived memory semantics. Independently, the vector backend uses lane-level source-read completion to delimit source operand lifetimes and release pure write-after-read (WAR) dependences early after source data capture, forming source-lifetime-aware scheduling. Software provides only packing permissions and static safety constraints, while vector dependences and request progression remain dynamically managed by the backend. This division of responsibilities strengthens coordination among instruction supply, memory access, and vector execution without changing the architectural semantics of standard RVV instructions.

The main contributions of this work are as follows:

\begin{itemize}

\item A hybrid-decoupled execution architecture and EP formation mechanism for RVV. The design integrates task-level decoupling, local instruction supply, HINT-guided EP formation, and hybrid dispatch to organize an interleaved scalar/vector instruction stream into a continuous EP stream, thereby alleviating instruction supply and scalar-side progress constraints.

\item A mechanism for propagating packet-derived prefetch intent to individual backend requests. The design keeps prefetch intent bound to the corresponding load throughout EP formation, hybrid dispatch, and request serialization and generates request-bound prefetch context, enabling greater overlap between memory access and vector computation.

\item A source operand-lifetime-aware vector scheduling mechanism. Lane-level source-read completion delimits source operand lifetimes and enables the early release of pure WAR dependences, while RAW, WAW, and mixed dependences remain under conventional management, reducing conservative waits while preserving correctness.

\item A synthesizable RTL implementation and systematic evaluation. We implement SEAM-V in synthesizable RTL and evaluate its performance benefits and hardware costs through cycle-accurate simulation, ablation and vector-length sensitivity analyses, microarchitectural counters, synthesis, power analysis, and physical design assessment.

\end{itemize}

The remainder of this paper is organized as follows. Section~II introduces the background and related work. Sections~III and~IV present the SEAM-V hardware architecture and software interface, respectively. Section~V reports the performance and hardware cost evaluation. Section~VI concludes the paper.

\section{Background and Related Work}
\label{sec:background}

This section introduces the spatiotemporal execution characteristics of RVV and the instruction supply constraints of tightly coupled organizations, reviews related execution models, and identifies the backend visibility gap of packet-derived memory intent and the limited use of source operand lifetime information. It then positions SEAM-V within this design space.

\subsection{RVV Spatiotemporal Execution and Scalar-Driven Instruction-Supply Constraints}
\label{sec:rvv_execution_supply}

SIMD and vector architectures both exploit data-level parallelism, but organize it differently. In a fixed-width SIMD architecture, the number of elements processed by an instruction is determined primarily by the ISA-visible register width and element type and is therefore largely fixed at the architectural level. In contrast, a vector architecture decouples architectural vector length from physical datapath width, allowing a vector instruction to expand into a stream of element operations according to runtime state and to progress across multiple lanes and beats. In RVV, \texttt{vl}, \texttt{SEW}, and \texttt{LMUL} determine the number of active elements, element width, and register grouping, respectively, while the number of physical lanes and available backend resources determine the spatiotemporal expansion of the element operations. Masking restricts the elements that update architectural state, and vector memory operations are further affected by the access mode, address generation, and memory system state. The vector backend is therefore not merely a fixed-width SIMD datapath, but a stateful execution system that must manage register dependences, execution resources, memory ordering, and instruction progress according to runtime state. Consequently, the actual timing of vector request progression cannot be determined statically by the front end~\cite{risc_v_vector_extension_1_0,arm_scalable_vector_extension}.

Many existing RVV processors adopt a tightly coupled organization in which the scalar core performs instruction fetch and decode, advances control flow, maintains address and loop state, updates vector configuration, and submits requests to the vector unit in program order. Vector-request generation depends on timely updates to the corresponding scalar state; branch resolution, address calculation, loop progression, and \texttt{vset} configuration may therefore enter the critical supply path of subsequent requests and introduce gaps between them. The resulting limitation is not simply insufficient front-end bandwidth, but the tight coupling among scalar state progression, request generation, and instruction-at-a-time delivery. When vector lengths are short or scalar and vector operations are frequently interleaved, fixed control and delivery overheads are difficult to amortize, and supply gaps more readily reduce lane and memory unit utilization. Even at larger vector lengths, kernels with frequent scalar state updates may remain constrained by the progression rate of the scalar path. The performance of tightly coupled RVV processors is therefore often limited by scalar path constraints on vector request supply rather than by insufficient data-parallel capacity in the vector backend~\cite{ara2_exploring_single_multi_core_vector_processing}.

\subsection{Related Execution Models and Backend Coordination Challenges}
\label{sec:related_models}

Existing research has enhanced backend execution capability through scalable vector datapaths and dynamic scheduling. The Ara family adopts a scalable lane-based organization to support long-vector execution, and Ara2 further shows that short-vector performance can be constrained by the issue rate of the scalar core~\cite{ara_1ghz_scalable_energy_efficient_risc_v_vector_processor,ara2_exploring_single_multi_core_vector_processing}. Espasa et al.\ introduced register renaming and out-of-order issue into vector processors to improve functional unit utilization and hide long-latency events~\cite{out_of_order_vector_architectures}. CODE combines a clustered vector register file, renaming, and execution decoupling to improve register file scalability while tolerating cross-cluster transfer and memory latency~\cite{overcoming_limitations_conventional_vector_processors}. For RVV, Saturn uses fine-grained chaining, multi-issue out-of-order scheduling, and run-ahead memory access to improve short-vector execution efficiency~\cite{instruction_scheduling_saturn_vector_unit}. Titan-I combines lane-based out-of-order execution, issue-as-commit, fine-grained chaining, and memory interleaving to jointly improve instruction-level, data-level, and memory-level parallelism~\cite{titan_i_open_source_high_performance_risc_v_vector_core}. These works primarily improve vector backend scalability, resource utilization, and tolerance to long-latency events.

Another line of work reduces the sustained constraint imposed by control paths on data-parallel execution by decoupling instruction supply, computation, and memory access. The vector-thread model of Scale uses vector-fetch commands and local instruction buffering to amortize control and supply overheads~\cite{implementing_scale_vector_thread_processor}. Hwacha adopts decoupled vector fetch to separate vector instruction acquisition and execution from the scalar control path~\cite{hwacha_vector_fetch_architecture_manual}. Stream Semantic Registers implicitly trigger regular memory accesses through stream-semantic registers, reducing explicit load/store instructions and their associated supply overheads~\cite{stream_semantic_registers}. Snitch's FREP mechanism repeatedly supplies floating-point instructions from a micro-loop buffer, reducing the need for the integer control pipeline to continuously drive the floating-point pipeline~\cite{snitch_tiny_pseudo_dual_issue_processor}. Decoupled Access/Execute separates access and execution streams through queues, allowing address generation and memory access to progress ahead of computation~\cite{decoupled_access_execute_computer_architectures}. Decoupled Vector Runahead further speculatively vectorizes indirect memory access chains to generate future requests and hide irregular memory latency~\cite{decoupled_vector_runahead}. These designs mainly improve sustained instruction supply and the overlap between memory access and computation, allowing requests to be generated earlier and to progress more continuously.

Complementary to these dynamic-execution and continuous-supply mechanisms, VLIW/EPIC and several DSP architectures use compile-time scheduling and ISA encoding to organize local operations and their issue relationships explicitly as instruction bundles or execution packets~\cite{fisher_trace_scheduling,fisher_eli512_vliw}. SLAP further relaxes the lockstep constraint among conventional VLIW functional units and supports SIMD vector lengths that vary at runtime~\cite{slap_split_latency_adaptive_vliw_pipeline_architecture}. These approaches preserve local instruction organization information at the front end, but RVV instructions must still be dynamically expanded in the backend according to runtime vector state and physical resources. Execution packets therefore cannot statically determine the actual progression of individual vector requests. When decoupled supply and explicit execution packets are introduced into RVV, the front end organizes multiple instructions at packet granularity, whereas the vector backend schedules and executes individual requests. After dispatch and serialization, packet-derived memory intent cannot participate in backend memory decisions unless the request interface preserves the hint associated with the corresponding load. Meanwhile, conventional dependence management makes limited use of the actual runtime consumption state of source operands and may retain pure WAR constraints longer than necessary. The former represents a backend visibility gap for packet-derived memory semantics during the conversion in execution granularity, while the latter reflects insufficient use of runtime source operand lifetime information. This work therefore does not seek to replace dynamic backend management with execution packets, but instead connects EP organization, request serialization, and backend execution while preserving memory intent and exploiting actual source operand lifetimes to reduce unnecessary dependence waits.

\subsection{Positioning of SEAM-V}
\label{sec:positioning}

SEAM-V consists of a hybrid-decoupled execution architecture and two key backend coordination mechanisms. The hybrid-decoupled path uses EPs as the basic unit for organizing and handing off scalar/vector instructions, while vector dependence management and request progression remain governed by runtime state in the backend; the two coordination mechanisms strengthen the connection between front-end instruction organization and dynamic backend execution without changing this division of responsibilities. This paper studies the complete execution path comprising task-level decoupling, HINT-guided EP formation, hybrid dispatch, backend-visible packet semantics, and source-lifetime-aware scheduling. Compared with our prior work~\cite{boosting_vector_instruction_throughput_risc_v_hybrid_decoupled_vliw}, SEAM-V further preserves prefetch intent throughout EP formation, request serialization, and backend delivery, integrates request-bound prefetching into the vector memory path, and uses lane-level source-read completion to determine source operand lifetimes and release pure WAR dependences early. The evaluation additionally includes ablation and sensitivity analyses, synthesis, power analysis, and physical design assessment to quantify the performance, energy-efficiency, and hardware costs of the key mechanisms.

\section{SEAM-V Hardware Architecture}
\label{sec:architecture}

This section presents the SEAM-V hardware architecture, focusing on EP formation and hybrid dispatch, the propagation of backend-visible memory semantics, and the early release of pure write-after-read (WAR) dependences based on source operand lifetimes.

\subsection{Architecture Overview}
\label{sec:architecture_overview}

\begin{figure*}[t]
    \centering
    \includegraphics[width=0.80\textwidth]{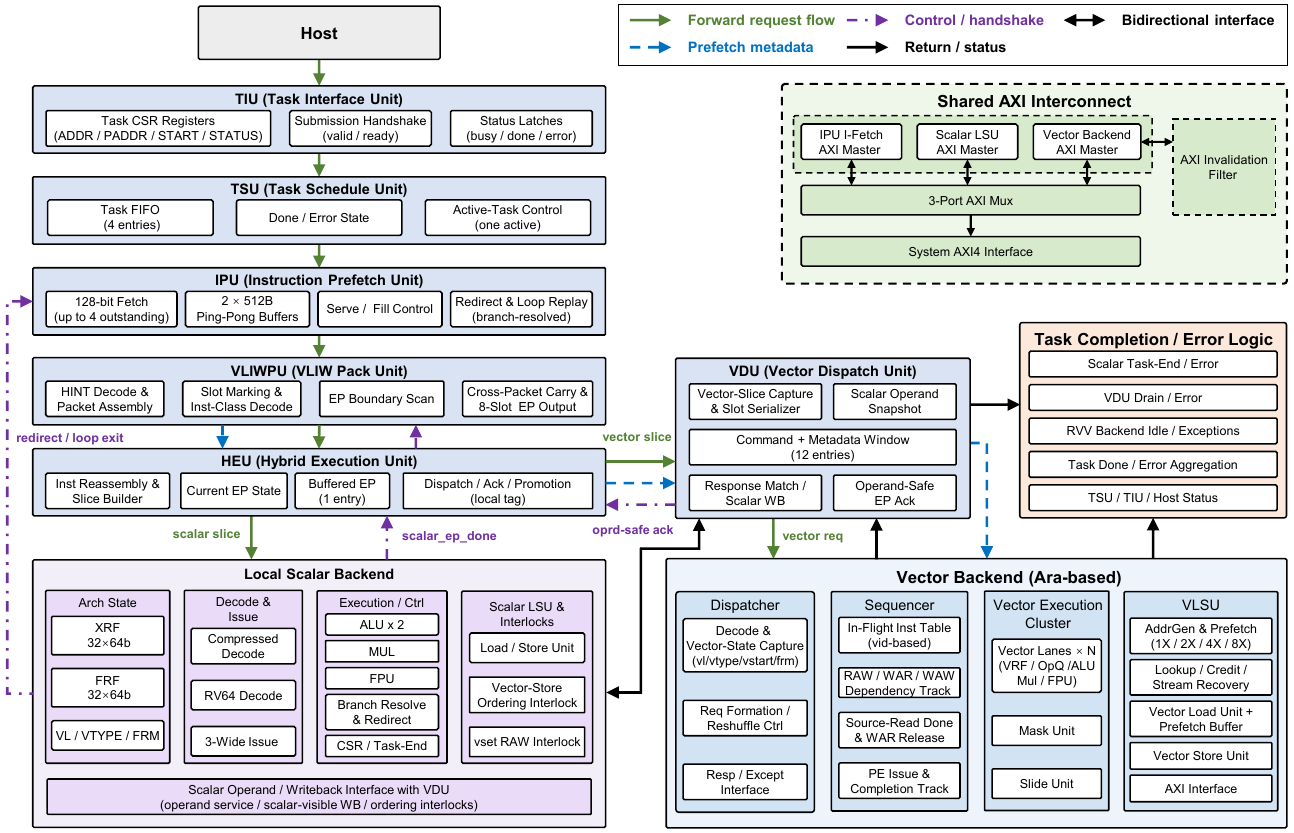}
    \caption{Overall microarchitecture of SEAM-V.}
    \label{fig:seamv_overview}
\end{figure*}

Fig.~\ref{fig:seamv_overview} shows the overall microarchitecture of SEAM-V. SEAM-V establishes a hybrid-decoupled execution path between the Host and the vector backend, connecting task submission and scheduling, local instruction supply and scalar execution, HINT-guided EP formation, hybrid dispatch, packet-derived memory semantics propagation, and source-lifetime-aware scheduling.

The Host submits a kernel entry address and execution context through the Task Interface Unit (TIU). The Task Schedule Unit (TSU), Instruction Prefetch Unit (IPU), VLIW Pack Unit (VLIWPU), Hybrid Execution Unit (HEU), Vector Dispatch Unit (VDU), local scalar backend, and vector backend jointly perform task scheduling, instruction supply, loop progression, and hybrid execution. The VLIWPU organizes the local instruction stream into EPs, and the HEU partitions each EP into a scalar slice and a vector slice. The scalar slice is delivered to the local scalar backend for task-local scalar computation, address updates, control flow, and memory operations. The VDU serializes the vector slice into individual instructions, captures the scalar operands required by each request, and delivers the resulting requests to the vector backend. The vector backend schedules and executes vector instructions and returns scalar-visible results to the local scalar backend through the VDU. Once the scalar and vector sides have completed a safe handoff, the front end may advance to subsequent EPs, while previously delivered vector requests may still be executing in the backend.

An EP is the basic unit of local instruction organization and scalar/vector handoff. The memory access intent carried by a HINT remains associated with the corresponding load throughout EP formation, request serialization, and buffering, and is used by the vector memory path as request-bound prefetch context. Packet-derived memory semantics therefore remain visible to the backend beyond the individual request interface. Independently, the vector backend uses source-read completion reported by the lanes to release pure WAR dependences after an older reader has safely captured its source data; RAW, WAW, and mixed dependences continue to be handled by the conventional mechanisms. The IPU, local scalar memory path, and vector memory path access system memory through a shared interconnect, while task completion and exception status are aggregated by unified logic and returned to the Host through the TSU and TIU.

\subsection{Task-Level Decoupling and Local Instruction Supply}
\label{sec:task_local_supply}

SEAM-V employs task-level decoupling to raise the interaction granularity between the Host and the vector coprocessor from individual vector instruction delivery to kernel-level task submission. The Host configures the task entry and descriptor through the TIU and triggers execution. The TSU buffers pending tasks in submission order and starts the task at the head of the queue when no task is active and the IPU is ready to accept it. Instruction fetch, loop progression, EP formation, and vector dispatch are subsequently performed within the coprocessor, without requiring the Host to participate continuously in instruction supply.

The IPU initializes its fetch and execution positions from the task entry, continuously fetches 128-bit instruction packets, and stores them in two parameterized ping-pong buffers, each configured as 512~B in the evaluated implementation. Once the first packet returns, the IPU can begin supplying the VLIWPU while continuing to fill the current buffer. After the current buffer has been filled, the other buffer fetches subsequent instructions in the background, thereby overlapping instruction consumption with memory access. The IPU maintains independent request and response indices and supports up to four ordered requests in flight in the evaluated configuration, limiting the occupation of the shared memory interface by instruction fetches. When the active buffer is exhausted, the IPU switches to a ready background buffer; if the background buffer is not yet ready, downstream instruction supply is paused. The IPU also reads ahead the next valid packet to hide the access latency of the on-chip instruction buffer.

For loop regions, the IPU combines the \texttt{loop\_start} and \texttt{loop\_end} markers in the HINT header with precise branch outcomes returned by the local scalar backend to implement buffer-resident replay. The \texttt{loop\_start} marker initiates loop-region construction, whereas \texttt{loop\_end} completes region locking and protects the corresponding buffers. After the packet containing \texttt{loop\_end} has been fetched, the IPU stops issuing new background fetch requests and accepts only responses for requests already in flight, thereby limiting further fetching beyond the loop body. For a taken backward branch, if the redirect target resides in a protected buffer, the IPU directly updates the execution position and switches buffer roles when necessary; instruction refetch is required only when the target is not resident. For a not-taken backward branch, the IPU releases loop protection and resumes the fall-through path. This mechanism reduces instruction fetch gaps during task startup and between loop iterations and continuously supplies the VLIWPU with fetch packets carrying accurate program counters.

\subsection{HINT-Guided Execute Packet Formation}
\label{sec:ep_formation}

The VLIWPU organizes the continuous fetch packets supplied by the IPU into EPs. Each logical packet begins with a 32-bit HINT header, which is decoded by the VLIWPU but is not itself packed into an EP; only the following instruction payload participates in EP formation. The \texttt{packet256} field determines whether two adjacent 128-bit fetch packets are combined into a 256-bit logical packet, whereas \texttt{pbits} and \texttt{cross} specify permission to continue packing within a logical packet and across logical packet boundaries, respectively. The remaining fields carry loop markers and memory access hints. Enlarging a logical packet increases only the observation window of the VLIWPU; an individual EP remains limited to at most eight 16-bit slots. The detailed encoding and software annotation rules are presented in Section~\ref{sec:software_ep_semantics}.

The VLIWPU scans sequentially from the current valid position of a logical packet using 16-bit slots as the basic granularity. A compressed instruction occupies one slot, whereas a 32-bit instruction occupies two consecutive slots~\cite{risc_v_instruction_set_manual_volume_i_unprivileged_architecture}. Hardware first identifies complete instruction boundaries, ensuring that an EP is never terminated in the middle of an instruction. If the \(p\)-bit associated with the boundary after the current instruction is set, the next instruction may continue to be packed into the same EP. The current EP ends when the corresponding \(p\)-bit is cleared, an explicit dependence boundary is detected, the eight-slot width limit is reached, or a branch or system instruction is encountered. A branch or system instruction may be the last instruction of an EP, but no subsequent instruction may be packed after it. Unconsumed instructions continue to form the next EP. At the end of a logical packet, if \texttt{cross} permits continuation, the EP has available slots, and no control boundary has been crossed, the VLIWPU buffers the tail fragment and continues packing from the next logical packet; otherwise, the EP terminates at the packet boundary. After entering the next logical packet, subsequent EP membership remains governed by that packet's \(p\)-bits, explicit dependence boundaries, control boundaries, and width constraints, preventing physical fetch boundaries from unnecessarily truncating an EP.

\begin{figure}[t]
    \centering
    \includegraphics[width=0.85\columnwidth]{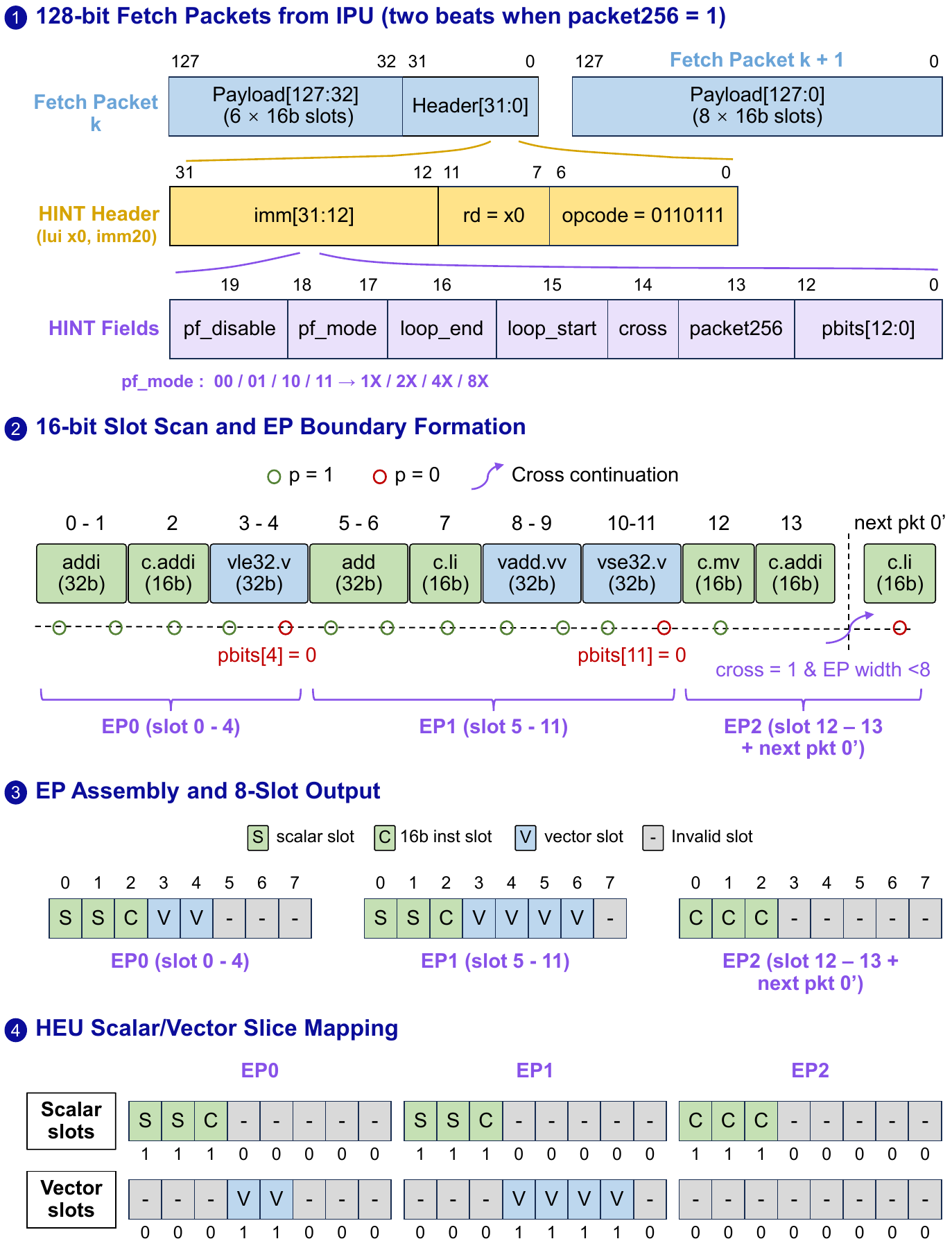}
    \caption{HINT-guided EP formation and scalar/vector slice mapping. A logical packet spans one or two 128-bit fetch packets, whereas each EP contains at most eight 16-bit slots. EP boundaries are determined by \(p\)-bit connectivity, cross-packet permission, and hardware constraints. The bottom panel shows per-slot validity in the scalar and vector slices.}
    \label{fig:ep_formation}
\end{figure}

Each formed EP carries per-slot valid bits, instruction length metadata, program counters, and instruction classes and is delivered to the HEU through a ready/valid interface. Downstream backpressure pauses EP transfer but does not alter membership that has already been determined. Prefetch intent also remains aligned with the corresponding instructions during ordinary packing and cross-packet buffering. Software-provided packing permissions and hardware-enforced boundaries jointly determine the final EP and provide the inputs required for subsequent scalar/vector slice formation and request-bound prefetch context generation. Cross-packet buffering also retains the prefetch fields of the initiating logical packet, so the resulting cross-packet EP preserves the prefetch context of the packet in which it begins.

\subsection{Hybrid Dispatch and EP Advancement}
\label{sec:hybrid_dispatch}

As shown at the bottom of Fig.~\ref{fig:ep_formation}, after an EP has been formed, the HEU uses the instruction length metadata to reconstruct complete instructions and partitions the EP into a scalar slice and a vector slice according to instruction class. The scalar slice is delivered to the local scalar backend for scalar computation, address updates, control flow, and memory operations. The vector slice enters the VDU, which processes its valid instructions in order, snapshots the required scalar operands, and forms the individual requests accepted by the vector backend. Prefetch intent remains associated with the corresponding load during serialization and buffering, while scalar-visible results are returned to the local scalar backend through the VDU. Because the two slices may be delivered in parallel, cross-path RAW dependences, scalar-visible-result dependences, and strict memory ordering requirements must be constrained by EP boundaries; the corresponding software rules are described in Section~\ref{sec:hint_encoding}. Dependences within the same execution path continue to be handled by the corresponding backend.

SEAM-V defines vector slice acknowledgement as an operand-safe handoff rather than completion of vector execution. The VDU acknowledges the vector-side handoff after all instructions in the vector slice have completed the required scalar operand snapshots and have been safely accepted by the VDU, and after the required scalar-visible results have returned. Operand snapshotting ensures that subsequent updates to the scalar register file cannot change the inputs of already formed requests, while the return of scalar-visible results ensures that subsequent scalar consumers observe the correct values. The HEU additionally waits for the scalar slice of the current EP to complete before confirming that both sides of the EP have completed their handoff. If a buffered EP is present, it is promoted to the current EP and the HEU reopens its input. At this point, previously delivered vector requests may still be executing in the backend.

To overlap the handoff of adjacent EPs, the HEU maintains one current EP and one buffered EP. A buffered scalar slice executes in EP order only after promotion. A buffered vector slice may be delivered early to the VDU only when the current EP contains no control-flow instruction, no scalar memory ordering operation remains active, no cross-EP GPR or FPR conflict recognizable by the HEU is present, and the VDU has sufficient capacity to accept the slice. Cross-EP vector register dependences continue to be maintained by the vector backend. This mechanism therefore provides only controlled vector-side early issue and does not constitute general cross-EP out-of-order execution. The command window within the VDU stores requests whose scalar operands have already been snapshotted, together with their required metadata and side effect information, absorbing short periods of backend backpressure while preserving request order. A task is declared complete only after the local scalar path reaches the task end and all vector-side buffers, request tracking state, and in-flight backend operations have drained.

\subsection{Backend-Visible Packet Semantics}
\label{sec:backend_visible_packet}

Prefetch intent expressed in a HINT remains associated with the corresponding load throughout EP formation, HEU delivery, and VDU serialization and is converted into request-bound prefetch context after the individual request is formed. We refer to this mechanism, in which packet-derived memory intent remains recognizable and usable by the vector backend after the conversion in execution granularity, as backend-visible packet semantics. The context carries prefetch valid state, explicit disable state, distance mode, and stream identity rather than acting as a global prefetch configuration for the vector backend. The 1X, 2X, 4X, and 8X modes set the prefetch distance to one, two, four, and eight times the logical access span of the current load, respectively, while the prefetched data length remains equal to that of the current access. For an eligible unit-stride load, the address generation unit computes a candidate prefetch address from the demand address, logical access span, and distance mode and checks the access type, remaining access range, and page boundary to avoid generating inapplicable or out-of-range prefetches.

A candidate prefetch must obtain address queue space, an address tracking lookup entry, and sufficient prefetch buffer credit before being issued through memory arbitration. Demand requests have higher priority; if the required conditions are not satisfied or resources are unavailable, the prefetch is not issued and the current load continues through the normal demand path. Returned prefetch data are stored in the VLSU prefetch buffer and associated with the corresponding address and coverage range. If a subsequent load hits the buffer and its required data are fully covered, it consumes the prefetched data and writes them to the vector register file through the normal load result path; otherwise, the access is completed through the conventional demand path. A prefetch return only fills the prefetch buffer, while architectural state is updated by the subsequent load through its normal execution path.

Issuing future data accesses early allows memory latency to overlap with current vector computation and other in-flight requests. The actual benefit depends on access stream regularity, prefetch distance, available memory-level parallelism, and queue, credit, and bandwidth pressure in the memory system~\cite{feedback_directed_prefetching,best_offset_hardware_prefetching,compiler_algorithm_prefetching}. A distance that is too short may fail to hide sufficient latency, whereas a distance that is too long may increase useless memory traffic and resource occupancy. If a subsequent demand address diverges from the tracked prefetch sequence, the vector memory path stops generating and issuing new prefetches and clears the invalid lookup entries and prefetch buffer data after the associated in-flight transactions have drained. When a prefetch is disabled, suppressed because of resource constraints, or misses, the load proceeds through the normal demand path without affecting architectural correctness.

\subsection{Source-Lifetime-Aware Scheduling}
\label{sec:source_lifetime}

Independently of the packet-derived prefetch mechanism, SEAM-V uses runtime state in the vector backend to optimize vector register dependence management. Conventional scheduling commonly retains a reader's association with its source vector registers until the entire instruction completes. Because a vector instruction may require multiple lanes and cycles to read its complete source data, its WAR dependences may remain active longer than necessary. In contrast, out-of-order scalar processors can generally eliminate WAR dependences through register renaming~\cite{tomasulo_multiple_arithmetic_units}, while operand reading latency in an in-order scalar pipeline is relatively short. Once all required source data have entered the operand queues within the lanes, the subsequent execution of the older reader no longer accesses the original vector registers. SEAM-V therefore uses source-read completion to delimit source operand lifetimes and release pure WAR dependences early, forming source-lifetime-aware scheduling.

When a vector instruction enters a lane, the lane sequencer records the operand requesters actually used by that instruction. Each requester reports completion after the complete source data stream has been written into the corresponding operand queue. Once all tracked source requests for the instruction have completed within a lane, that lane reports source-read completion to the main sequencer. The main sequencer aggregates the completion state across all lanes and marks the reader as source released after every lane has completed source data capture. Existing pure WAR dependences that point to the released reader can then be removed early, and subsequent requests that write the same vector register no longer establish new WAR constraints against that reader. The writer may therefore overlap with the later execution of the older read instruction.

This mechanism shortens only the lifetime of pure WAR dependences. RAW and WAW dependences, as well as mixed dependences that include another dependence type, continue to be maintained according to conventional completion events and therefore preserve producer--consumer relationships and architectural write ordering. Because all source data required by the older reader have already entered the operand queues, subsequently overwriting the original register cannot affect its execution result. \texttt{VRGATHER}, \texttt{VRGATHEREI16}, and \texttt{VCOMPRESS} use specialized source request mechanisms and continue to use conservative dependence handling in the current implementation. Source-lifetime-aware scheduling primarily benefits cases in which source data are read substantially earlier than full instruction completion; when no pure WAR constraint blocks progress, the original scheduling behavior remains unchanged.

\section{Software Interface and EP Semantics}
\label{sec:software_ep_semantics}

This section describes the SEAM-V programming model and EP semantics from the software perspective and explains how task-level execution, HINT encoding, EP formation rules, and safety constraints provide hardware with packing permissions, local loop markers, and memory access intent.

\subsection{Task-Level Programming Model}
\label{sec:task_programming_model}

SEAM-V uses a task as the basic unit of software submission and execution management. The main program is responsible for data preparation, parameter passing, and task submission, while performance-critical RVV kernels are encapsulated as SEAM-V tasks. Vector computation within a task remains expressed using standard RVV instructions, with HINT annotations providing the structural information and hints required for EP formation, local loop instruction supply, and request-bound prefetching.

Software responsibilities can be divided into two levels. At the algorithmic level, data-parallel computation remains expressed as an ordinary RVV kernel. At the execution structure level, software provides the packing permissions and control boundaries required for EP formation, the loop markers required for local replay by the IPU, and the memory access hints required for backend prefetching. Software does not explicitly manage vector backend state. Instead, within locally analyzable regions, it identifies which adjacent instructions may be organized into the same EP, where EP boundaries must be formed, and which regular memory streams are suitable for request-bound prefetching. SEAM-V thereby establishes a lightweight software--hardware contract on top of the standard RVV programming model, allowing hardware to use software-known local structure and memory access intent without reconstructing this information after the instruction stream has been converted into individual requests.

\subsection{HINT Encoding and Software Constraints}
\label{sec:hint_encoding}

SEAM-V uses the RISC-V HINT form \texttt{lui x0, imm20} to encode the packet-level information required for EP formation, local loop instruction supply, and request-bound prefetching. The HINT does not modify architecturally visible state~\cite{risc_v_instruction_set_manual_volume_i_unprivileged_architecture}. The VLIWPU parses its immediate field and packs only the following payload; the HINT header itself does not enter an EP. Table~\ref{tab:hint_header} summarizes the encoding.

\begin{table}[t]
    \centering
    \caption{SEAM-V HINT header encoding.}
    \label{tab:hint_header}
    \setlength{\tabcolsep}{2.3pt}
    \renewcommand{\arraystretch}{1.08}
    \begin{tabularx}{\columnwidth}{
        @{}
        l
        l
        >{\raggedright\arraybackslash}X
        @{}
    }
        \toprule
        Field & Bits & Meaning \\
        \midrule

        \texttt{pbits} &
        \texttt{imm20[12:0]} &
        Allows packing across selected 16-bit slot boundaries. \\

        \texttt{packet256} &
        \texttt{imm20[13]} &
        Extends the logical packet to 256 bits. \\

        \texttt{cross} &
        \texttt{imm20[14]} &
        Allows cross-packet EP continuation. \\

        \texttt{loop\_start} &
        \texttt{imm20[15]} &
        Marks the start of an IPU-local loop. \\

        \texttt{loop\_end} &
        \texttt{imm20[16]} &
        Marks the end of an IPU-local loop. \\

        \texttt{pf\_mode} &
        \texttt{imm20[18:17]} &
        Selects the 1X, 2X, 4X, or 8X prefetch distance. \\

        \texttt{pf\_disable} &
        \texttt{imm20[19]} &
        Disables prefetching. \\

        \bottomrule
    \end{tabularx}
\end{table}

The HINT header defines the software-visible annotations associated with a logical packet. The \texttt{packet256} field enlarges the local scanning and packing window of the VLIWPU, while \texttt{cross} permits an unfinished EP at the end of a logical packet to continue into the next logical packet when the required conditions are satisfied. Neither field changes the maximum width of an individual EP. The \texttt{loop\_start} and \texttt{loop\_end} fields identify a local loop region that can be cached and replayed by the IPU. Actual loop progression remains determined by the backward branch resolved by the local scalar path, and the two fields are not propagated with vector requests. A cross-packet EP retains the prefetch context of the logical packet in which it begins; consequently, the prefix incorporated from the next logical packet must not contain a vector load that requires a different prefetch hint.

The \texttt{pbits} field describes permission to continue packing within a logical packet. When the \(p\)-bit associated with a slot boundary is set, the following instruction may continue to be packed into the current EP; when the bit is cleared, an EP boundary is formed. The VLIWPU reconstructs a 32-bit instruction from two consecutive 16-bit slots, so an EP boundary never splits an instruction. Branch and system instructions and the hardware width limit force the current EP to terminate. Scalar--scalar dependences within an EP are maintained in program order by the local scalar backend, while vector--vector RAW, WAR, and WAW dependences remain governed by the conventional hazard mechanisms in the vector backend. The \texttt{pbits} field therefore expresses only permission to continue packing and does not provide a dependence exemption.

Because scalar and vector slices may be delivered in parallel, software must establish EP boundaries for cross-path RAW dependences in which a scalar instruction writes a value read by a vector instruction, for dependences in which a vector instruction produces a scalar-visible result consumed by a scalar instruction, and for scalar/vector memory operations requiring strict program order. The current implementation provides a dedicated interlock for \texttt{vset} results; other scalar-visible vector results should be separated from their scalar consumers. An EP boundary only prevents parallel delivery within the same EP and does not guarantee that an earlier vector load or store has completed. Cross-path memory ordering therefore remains enforced by the scalar memory completion and in-flight vector store tracking mechanisms. When a vector instruction reads the current scalar value and a following scalar instruction updates iteration state, the updated value must be used only by later EPs. Software should conservatively insert an EP boundary when the dependence cannot be determined reliably.

Prefetching is jointly controlled by \texttt{pf\_disable} and \texttt{pf\_mode}. The former disables prefetching, whereas the latter selects a 1X, 2X, 4X, or 8X distance when prefetching is enabled. Prefetch hints are encoded at logical packet granularity. Hardware preserves their association with the corresponding requests throughout EP formation and request serialization, and an eligible unit-stride vector load triggers request-bound prefetching. Prefetch distance is measured in units of the logical access span of the current request: 1X targets the next equally sized data block, while 2X, 4X, and 8X progressively increase the distance. Software should disable the hint for gather/scatter operations, irregular accesses, short execution phases, or memory access patterns unsuitable for fixed-distance prefetching. These fields provide only the hints required for EP formation, local loop instruction supply, and memory access optimization and do not alter the architectural semantics of RVV instructions. The vector backend continues to maintain register dependences according to its conventional rules, and a vector load for which prefetching is not enabled continues through the normal demand path.

\subsection{RVV Kernel Example}
\label{sec:rvv_kernel_example}

Listing~\ref{lst:hint_kernel} uses a simplified RVV kernel to illustrate the relationship between the HINT fields and EP formation. The \texttt{HINT} directive is a pseudo-macro that expands to the corresponding RISC-V encoding.

\begin{figure}[t]
    \vspace{-1mm}
    \centering
    \refstepcounter{listing}
    \label{lst:hint_kernel}
    \setlength{\tabcolsep}{2.0pt}
    \renewcommand{\arraystretch}{0.82}
    \begin{tabular}{
        @{}
        >{\raggedright\arraybackslash}p{0.20\columnwidth}
        >{\raggedright\arraybackslash}p{0.75\columnwidth}
        @{}
    }
        \toprule
        \multicolumn{2}{@{}l}{Listing~\thelisting: RVV kernel with HINT-guided EP formation} \\
        \midrule
        \addlinespace[0.5pt]

        \asmcode{loop:} & \\

        \asmcode{\asmindent HINT} &
        \asmcode{pbits=0x08A8, packet256=1, cross=1,} \\
        &
        \asmcode{loop\_start=1, loop\_end=0,} \\
        &
        \asmcode{pf\_disable=0, pf\_mode=00} \\

        \addlinespace[1pt]

        \asmcode{\asmindent vsetvli} &
        \asmcode{t0, a0, e32, m1, ta, ma} \\

        \addlinespace[1pt]

        \asmcode{\asmindent vle32.v} &
        \asmcode{v0, (a1)} \\
        \asmcode{\asmindent vle32.v} &
        \asmcode{v1, (a2)} \\
        \asmcode{\asmindent vfadd.vv} &
        \asmcode{v2, v0, v1} \\
        \asmcode{\asmindent vse32.v} &
        \asmcode{v2, (a3)} \\

        \addlinespace[1pt]

        \asmcode{\asmindent sub} &
        \asmcode{a0, a0, t0} \\
        \asmcode{\asmindent slli} &
        \asmcode{t1, t0, 2} \\

        \addlinespace[2pt]

        \asmcode{\asmindent HINT} &
        \asmcode{pbits=0x000A, packet256=0, cross=0,} \\
        &
        \asmcode{loop\_start=0, loop\_end=0,} \\
        &
        \asmcode{pf\_disable=1, pf\_mode=00} \\

        \addlinespace[1pt]

        \asmcode{\asmindent add} &
        \asmcode{a1, a1, t1} \\
        \asmcode{\asmindent add} &
        \asmcode{a2, a2, t1} \\
        \asmcode{\asmindent add} &
        \asmcode{a3, a3, t1} \\

        \addlinespace[2pt]

        \asmcode{\asmindent HINT} &
        \asmcode{pbits=0x0000, packet256=0, cross=0,} \\
        &
        \asmcode{loop\_start=0, loop\_end=1,} \\
        &
        \asmcode{pf\_disable=1, pf\_mode=00} \\

        \addlinespace[1pt]

        \asmcode{\asmindent bnez} &
        \asmcode{a0, loop} \\
        \asmcode{\asmindent ret} & \\
        \asmcode{\asmindent nop} & \\

        \addlinespace[0.5pt]
        \bottomrule
    \end{tabular}
    \vspace{-1mm}
\end{figure}

The first HINT sets \texttt{packet256=1} and \texttt{cross=1} to enlarge the observation window of the logical packet and permit cross-packet EP formation. The \texttt{loop\_start} field marks the beginning of the local loop region, while \texttt{pf\_disable=0} and \texttt{pf\_mode=00} provide 1X prefetch intent for eligible unit-stride vector loads. The \texttt{pbits} fields and hardware-enforced conditions jointly determine the EP boundaries. In this example, the two loads, \texttt{vfadd.vv}, and \texttt{vse32.v} form one EP, while their internal RAW dependences remain handled by the vector backend. The \texttt{sub} and \texttt{slli} instructions form a cross-packet EP with \texttt{add a1} and \texttt{add a2} from the next logical packet; \texttt{add a3} enters the following EP because of the width limit.

The \texttt{vsetvli} instruction forms a separate EP, and its result in \texttt{t0} is made visible before subsequent reads through the existing result return and interlock mechanisms. The VDU snapshots the scalar operands \texttt{a1}, \texttt{a2}, and \texttt{a3} used by the vector memory instructions before the following scalar address updates. The cross-packet EP retains the prefetch context of the first HINT, and the incorporated prefix of the next logical packet contains only scalar address updates, introducing no conflicting prefetch hint. The second HINT disables prefetching for subsequent EPs formed from that logical packet. The third HINT marks the end of the local loop region through \texttt{loop\_end=1}; actual loop progression remains determined by the runtime outcome of \texttt{bnez}, and the control-flow instruction is forced by hardware to terminate the current EP. The example shows how HINT headers provide packing permissions, cross-packet continuation, loop markers, and memory access intent, allowing the VLIWPU to form EPs, the IPU to support local replay, and the vector memory path to perform request-bound prefetching.

\section{Experimental Methodology and Evaluation}
\label{sec:evaluation}

\subsection{Experimental Setup and Workloads}
\label{sec:experimental_setup}

We use cycle-accurate RTL simulation to collect execution cycles and microarchitectural counters. The six one-dimensional kernels sweep application vector length (AVL), whereas the remaining kernels use the fixed input sizes and execution configurations. The reference design is an Ara-based tightly coupled vector processor, referred to as the tightly coupled baseline (TC)~\cite{ara_1ghz_scalable_energy_efficient_risc_v_vector_processor,ara2_exploring_single_multi_core_vector_processing}. SEAM-V and TC execute identical workloads and input data with the same vector backend parameters, memory system configuration, clock frequency, and task measurement boundaries. Normalized performance is calculated from paired execution cycle measurements for each workload configuration. Table~\ref{tab:eval_config} summarizes the shared vector backend and memory system parameters, together with the local instruction supply and scalar execution resources introduced by SEAM-V. The ablation experiments use the same environment. Starting from the base hybrid-decoupled configuration, request-bound prefetching and source-lifetime-aware scheduling are enabled either individually or together, and the resulting configurations are compared with the full SEAM-V design to isolate the incremental contribution of each mechanism. Synthesis, power, and placement results are used to evaluate the area, power, and energy-efficiency costs of the added mechanisms.

\begin{table}[t]
    \centering
    \caption{Key hardware configuration.}
    \label{tab:eval_config}
    \scriptsize
    \setlength{\tabcolsep}{5pt}
    \renewcommand{\arraystretch}{1.08}
    \begin{tabular}{@{}ll@{}}
        \toprule
        Item & Configuration \\
        \midrule
        EP organization
            & 128-bit fetch, 8$\times$16-bit slots \\
        Local buffers
            & IPU: $2\times512$~B; VDU: 12 entries \\
        Backend capacity
            & 8 in-flight vector instructions \\
        Vector backend
            & 4 lanes, VLEN/ELEN = 1024/64 bits \\
        Local scalar backend
            & 3-wide issue, 2 simple ALUs \\
        Memory system
            & 128-bit AXI, 1~MiB L2 \\
        \bottomrule
    \end{tabular}
\end{table}

The workloads are divided into three groups~\cite{updated_blas,polybench_c,rodinia_benchmark_suite}. The first group consists of one-dimensional kernels with variable AVL, covering basic streaming operations, reduction, data exchange, filtering, transforms, and one-dimensional neighborhood computation. These kernels are used to analyze how vector length affects instruction supply efficiency and task-level cycles per processed element. The second group consists of BLAS and matrix kernels and is used to examine the effects of LMUL, blocking strategy, data reuse, and local instruction organization in compute-intensive workloads. The third group consists of fixed-size representative application kernels, covering two-dimensional convolution, two-dimensional iterative computation, softmax, molecular dynamics, and matrix--vector multiplication. These kernels are used to evaluate SEAM-V under execution patterns that more closely resemble application-level computation. Together, the three groups form the representative workload set and execution configurations summarized in Table~\ref{tab:workload_config}.

\begin{table}[t]
    \centering
    \caption{Representative workloads and configurations.}
    \label{tab:workload_config}
    \scriptsize
    \setlength{\tabcolsep}{1.7pt}
    \renewcommand{\arraystretch}{0.98}
    \begin{tabularx}{\columnwidth}{
        @{}
        >{\raggedright\arraybackslash}p{0.145\columnwidth}
        >{\centering\arraybackslash}p{0.06\columnwidth}
        >{\centering\arraybackslash}p{0.42\columnwidth}
        >{\centering\arraybackslash}X
        @{}
    }
        \toprule
        Kernel & Type & Operation & Configuration \\
        \midrule

        AXPY
        & \multirow[c]{6}{*}{FP32}
        & \(\mathbf{y}\leftarrow\alpha\mathbf{x}+\mathbf{y}\)
        & \multirow[c]{6}{*}{\shortstack[c]{\(\mathrm{AVL}\in\mathcal{A}\)\\rep. \(1024\); \texttt{m1}}} \\

        DOTP
        & & \(s\leftarrow\mathbf{x}^{\mathsf T}\mathbf{y}\) & \\

        SWAP
        & & \(\mathbf{x}\leftrightarrow\mathbf{y}\) & \\

        DWT
        & & \((l_i,h_i)\leftarrow(x_{2i}\pm x_{2i+1})/\sqrt{2}\) & \\

        STENCIL3
        & & \(y_i\leftarrow\sum_{k=-1}^{1}c_kx_{i+k}\) & \\

        FIR5
        & & \(y_i\leftarrow\sum_{k=0}^{4}h_kx_{i-k}\) & \\

        \midrule

        TRSM
        & \multirow[c]{6}{*}{FP32}
        & \(LX=B\)
        & \(N=128;\ \texttt{m4}\) \\

        GER
        & & \(A\leftarrow A+\alpha\mathbf{x}\mathbf{y}^{\mathsf T}\)
        & \(128^2;\ \texttt{m1}\) \\

        SYMV
        & & \(\mathbf{y}\leftarrow\alpha A_{\mathrm{sym}}\mathbf{x}+\beta\mathbf{y}\)
        & \multirow[c]{2}{*}{\(N=128;\ G=8;\ \texttt{m4}\)} \\

        GEMV
        & & \(\mathbf{y}\leftarrow\alpha A\mathbf{x}+\beta\mathbf{y}\)
        & \\

        GEMM
        & & \(C\leftarrow\alpha AB+\beta C\)
        & \(64^3;\ 4\text{-row};\ \texttt{m1}\) \\

        SYRK
        & & \(C\leftarrow\alpha AA^{\mathsf T}+\beta C\)
        & \(N=64;\ \texttt{m4}\) \\

        \midrule

        JACOBI2D
        & \multirow[c]{2}{*}{FP64}
        & \(B_{i,j}\leftarrow\frac{1}{5}\sum_{(p,q)\in\mathcal{N}_5(i,j)}A_{p,q}\)
        & \(128\times64;\ \texttt{m4}\) \\

        CONV2D
        & & \(Y\leftarrow X\ast K\)
        & \(64\times32;\ 3\times3;\ \texttt{m2}\) \\

        SPMV
        & \multirow[c]{3}{*}{FP32}
        & \(\mathbf{y}\leftarrow A\mathbf{x}\)
        & \(32\times32;\ 32~\mathrm{nnz/row};\ \texttt{m1}\) \\

        LAVAMD
        & & \(\mathbf{F}_i\leftarrow\sum_j\mathbf{f}_{ij}\)
        & \(N_p=256;\ \texttt{m1}\) \\

        SOFTMAX
        & & \(y_i\leftarrow e^{x_i}/\sum_j e^{x_j}\)
        & \(3\times256;\ \texttt{m1}\) \\

        \bottomrule
    \end{tabularx}

    \vspace{1pt}
    \parbox{\columnwidth}{
        \(\mathcal{A}=\{2^k\mid5\leq k\leq12\}\).
        The six 1D kernels sweep AVL over \(\mathcal{A}\), with 1024 used as the representative point;
        all other kernels use the listed fixed configurations.
        \(G\) denotes the number of row groups, and ``4-row'' denotes GEMM register blocking.
    }
\end{table}

\subsection{Overall Performance}
\label{sec:eval_overall}

\begin{figure}[t]
    \centering
    \includegraphics[width=\linewidth]{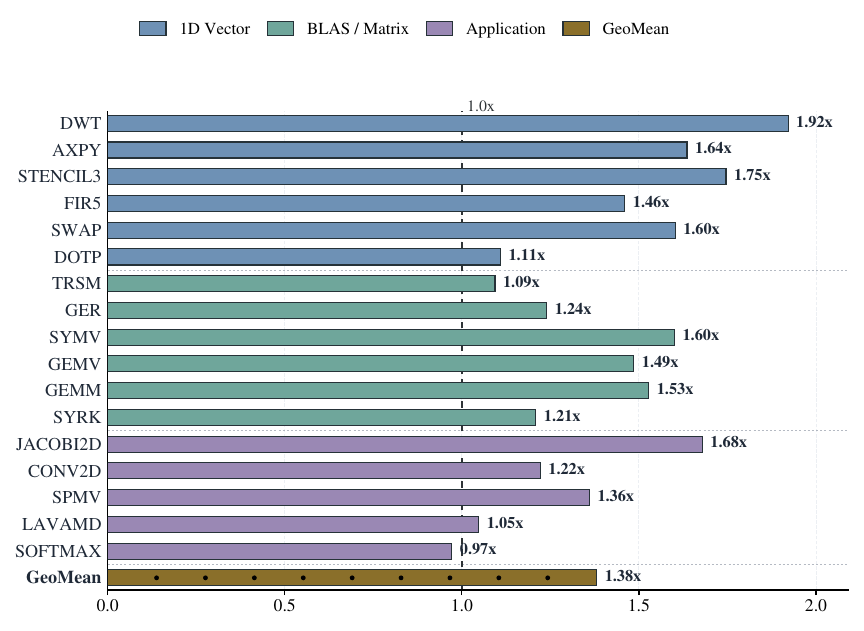}
    \caption{Representative speedup of the full SEAM-V design over the Ara-based tightly coupled baseline (TC).}
    \label{fig:overall_speedup}
\end{figure}

Fig.~\ref{fig:overall_speedup} reports the representative performance of the full SEAM-V design relative to TC. Speedup is defined as the task execution cycles of TC divided by those of SEAM-V. Across 17 representative kernel configurations, SEAM-V achieves a geometric mean speedup of approximately \(1.38\times\). The one-dimensional vector, BLAS and matrix, and fixed-size application groups achieve geometric mean speedups of approximately \(1.56\times\), \(1.35\times\), and \(1.23\times\), respectively. All representative points except SOFTMAX show positive gains, indicating that SEAM-V improves performance across a range of computation patterns, although the magnitude of the benefit remains workload dependent.

Among the one-dimensional vector kernels, DWT and STENCIL3 achieve approximately \(1.92\times\) and \(1.75\times\) speedup, respectively. AXPY and SWAP both exceed \(1.60\times\), FIR5 reaches approximately \(1.46\times\), and DOTP obtains a comparatively limited gain. This group achieves the highest geometric mean speedup, showing that SEAM-V broadly improves the sustained execution efficiency of one-dimensional vector kernels. The BLAS and matrix kernels also exhibit relatively stable improvements. SYMV, GEMM, and GEMV achieve approximately \(1.60\times\), \(1.53\times\), and \(1.49\times\) speedup, respectively; GER and SYRK obtain moderate gains, while TRSM reaches approximately \(1.09\times\). These results show that SEAM-V can reduce task execution cycles in compute-intensive matrix workloads, although the benefit depends on computation structure and execution configuration. Among the fixed-size application kernels, JACOBI2D achieves approximately \(1.68\times\) speedup, while SPMV and CONV2D reach approximately \(1.36\times\) and \(1.22\times\), respectively. LAVAMD shows a small improvement, whereas SOFTMAX reaches approximately \(0.97\times\) and exhibits a slight regression. SEAM-V can therefore extend its benefits to two-dimensional iterative computation, matrix--vector multiplication, and convolution, but not every workload converts the activity introduced by the additional mechanisms into a net performance gain.

Overall, all three workload groups achieve positive geometric mean improvements. The gains are largest for the one-dimensional vector kernels, followed by the BLAS and matrix kernels, while the fixed-size application group shows a more moderate overall benefit. These results indicate that the coordinated optimization of instruction supply, hybrid dispatch, and backend execution applies to multiple computation patterns, although the final benefit remains dependent on control structure, memory access behavior, and backend execution pressure.

\subsection{Vector Length Sensitivity and Per-Element Efficiency}
\label{sec:eval_avl_sensitivity}

\begin{figure*}[t]
    \centering
    \includegraphics[width=0.70\textwidth]{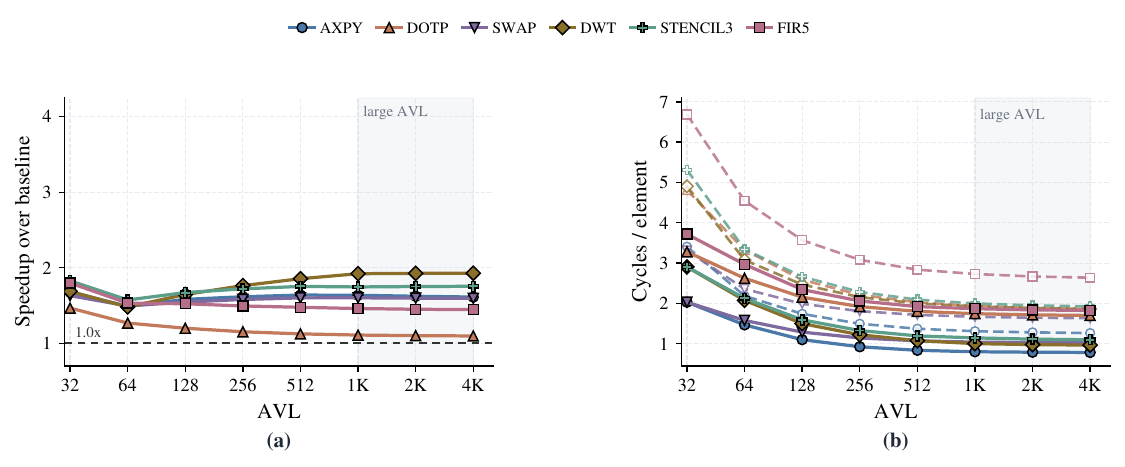}
    \caption{Vector-length sensitivity and per-element efficiency for six one-dimensional vector kernels. (a) Speedup of SEAM-V over TC. (b) Task-level cycles per element, where solid lines denote SEAM-V and dashed lines denote TC.}
    \label{fig:avl_sweep_cpe}
\end{figure*}

Fig.~\ref{fig:avl_sweep_cpe} further analyzes six one-dimensional vector kernels across different application vector lengths (AVLs). Fig.~\ref{fig:avl_sweep_cpe}(a) reports the speedup of SEAM-V over TC, while Fig.~\ref{fig:avl_sweep_cpe}(b) compares their cycles per element (CPE). CPE is defined as the task execution cycles divided by AVL and represents the average task-level cost of processing one element. This experiment examines how the performance benefit varies with vector granularity and distinguishes fixed overheads at short AVLs from per-element execution efficiency at large AVLs.

SEAM-V provides positive gains at every tested AVL. At AVL \(=32\), the geometric mean speedup across the six kernels is approximately \(1.68\times\). It decreases to approximately \(1.47\times\) at AVL \(=64\), indicating that SEAM-V's reduction of startup, control, and instruction supply overheads is more pronounced at short AVLs, while the relative contribution of these fixed-cost reductions is gradually amortized as more elements are processed. The speedup then increases and remains near \(1.55\times\) over the AVL \(=512\)--4096 range, indicating that the benefit at larger AVLs increasingly reflects improvements in sustained execution efficiency.

The speedup trends differ across kernels as AVL increases. DWT and STENCIL3 retain approximately \(1.93\times\) and \(1.75\times\) speedup at large AVLs, while AXPY and SWAP remain near \(1.60\times\). FIR5 gradually converges to approximately \(1.45\times\), whereas DOTP shows the smallest gain and decreases to approximately \(1.10\times\) at large AVLs, demonstrating that computation structure, dependence behavior, and memory access characteristics substantially affect the response to SEAM-V. The CPE curves further show that SEAM-V reduces task-level cycles per element at every tested AVL. As AVL increases, most kernels gradually enter a relatively stable CPE region. At AVL \(=4096\), the CPE values of AXPY, DWT, and FIR5 are approximately 0.79, 0.97, and 1.83 cycles per element, respectively. Taken together, the speedup and CPE results show that SEAM-V lowers per-element execution cost throughout the evaluated range, although the magnitude and evolution of the benefit remain kernel dependent.

\subsection{Ablation Study}
\label{sec:eval_ablation}

\begin{figure}[t]
    \centering
    \includegraphics[width=0.90\linewidth]{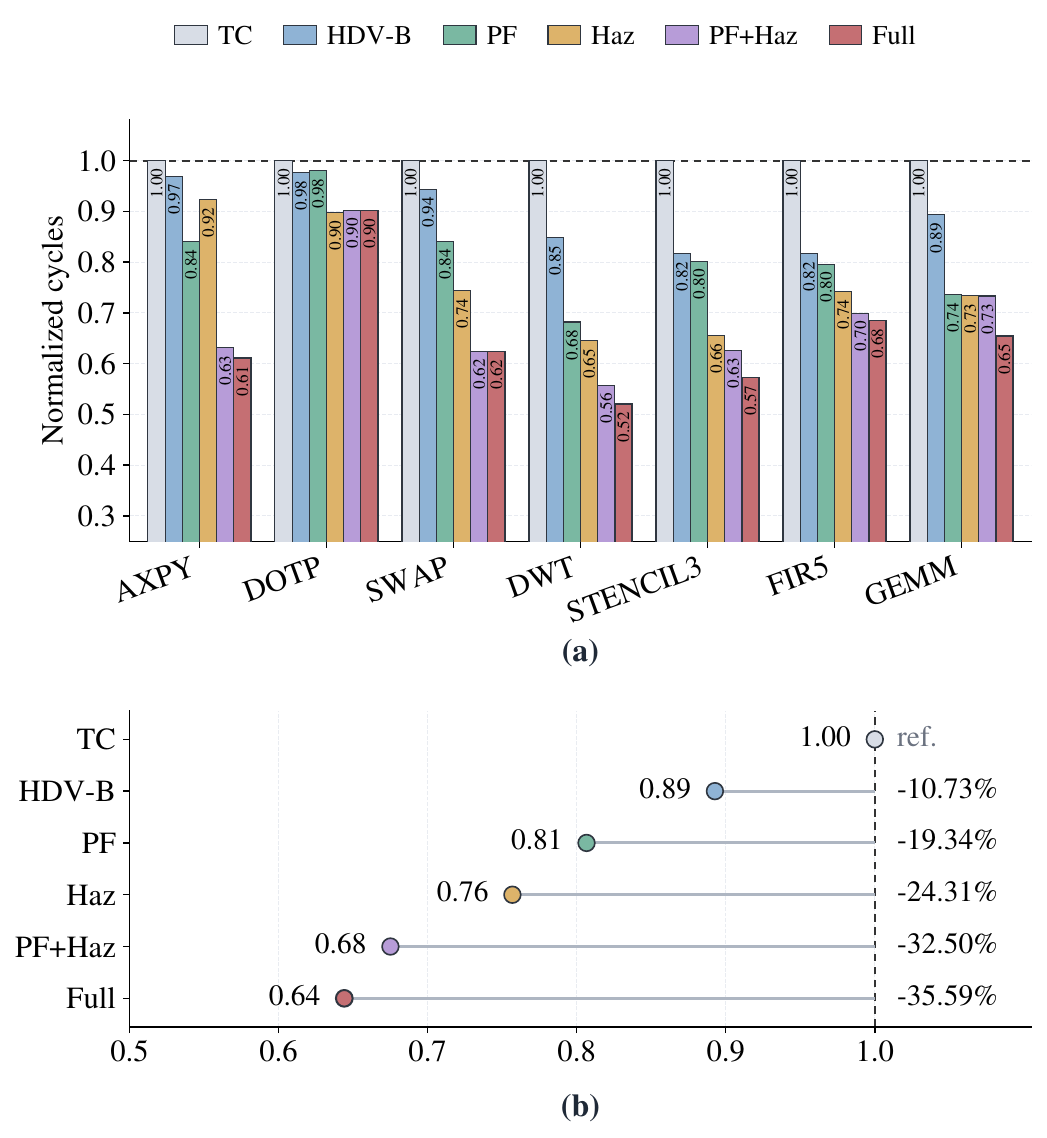}
    \caption{Ablation study of the key SEAM-V mechanisms. Task execution cycles are normalized to the tightly coupled baseline, with TC \(=1.0\); lower values indicate better performance.}
    \label{fig:ablation_study}
\end{figure}

Fig.~\ref{fig:ablation_study} evaluates the key SEAM-V mechanisms. Fig.~\ref{fig:ablation_study}(a) compares normalized task execution cycles across seven representative kernels, while Fig.~\ref{fig:ablation_study}(b) summarizes the geometric mean of each configuration. Normalized cycles are defined as the task execution cycles of each configuration divided by the corresponding TC cycles; lower values indicate better performance. We compare five SEAM-V configurations against TC. \texttt{HDV-B} is the base hybrid-decoupled configuration and includes task-level decoupling, local instruction supply, EP formation, and basic hybrid dispatch. \texttt{PF} and \texttt{Haz} enable request-bound prefetching and source-lifetime-aware scheduling on top of \texttt{HDV-B}, respectively; the latter uses source-read completion to release pure WAR dependences early. \texttt{PF+Haz} enables both mechanisms. \texttt{Full} is the complete SEAM-V implementation and further enables the full optimization path, including buffered vector early issue and scalar operand lookahead.

Across the seven representative kernels, \texttt{HDV-B} reduces the geometric mean normalized cycles to approximately 0.89, corresponding to a 10.73\% reduction relative to TC and showing that the base hybrid-decoupled execution path provides a clear performance benefit. \texttt{PF} and \texttt{Haz} further reduce the geometric mean to approximately 0.81 and 0.76, respectively, demonstrating that request-bound prefetching and source-lifetime-aware scheduling each provide an independent incremental contribution. Enabling both mechanisms in \texttt{PF+Haz} reduces the geometric mean to approximately 0.68, below either single-mechanism configuration, indicating that memory access overlap and early release of pure WAR dependences can jointly improve execution. The complete \texttt{Full} configuration reaches approximately 0.64, corresponding to a 35.59\% reduction relative to TC and a further reduction of approximately 4.58\% relative to \texttt{PF+Haz}.

The source of the improvement differs across kernels. AXPY benefits more from \texttt{PF} than from \texttt{Haz} when the two mechanisms are enabled separately, while their combination produces a more pronounced cycle reduction. DOTP, SWAP, STENCIL3, and FIR5 benefit more strongly from \texttt{Haz}; \texttt{PF} provides almost no additional improvement for DOTP and only limited standalone improvement for STENCIL3 and FIR5. DWT benefits from both mechanisms and continues to reduce its execution cycles under the combined and complete configurations. For GEMM, the standalone effects of \texttt{PF} and \texttt{Haz} are similar, and \texttt{PF+Haz} provides only a small additional gain over either single-mechanism configuration, although \texttt{Full} still obtains a further improvement. Overall, \texttt{HDV-B} provides the base hybrid-decoupling benefit, \texttt{PF} and \texttt{Haz} contribute incremental improvements through memory access overlap and dependence management, respectively, and \texttt{Full} captures their integrated effect together with the complete hybrid-dispatch optimization path.

\subsection{Microarchitectural Analysis}
\label{sec:eval_microarch}

\begin{figure*}[t]
    \centering
    \includegraphics[width=0.70\textwidth]{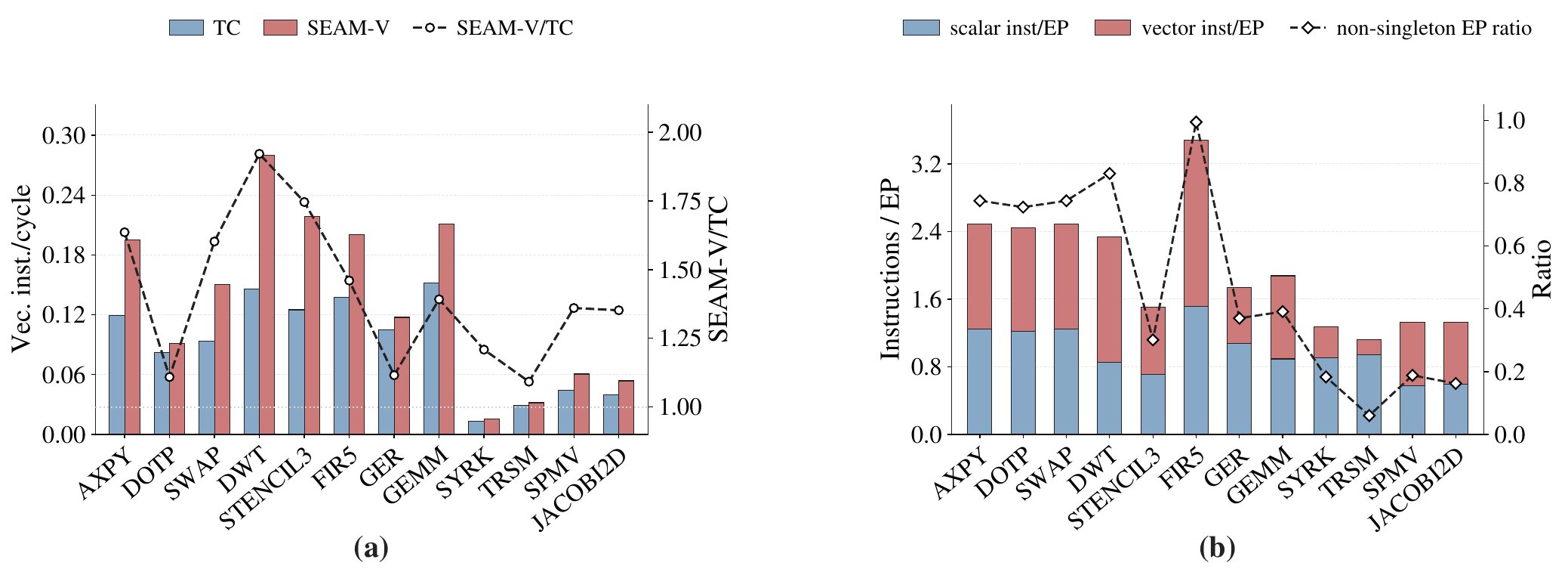}
    \caption{Instruction throughput and EP organization. (a) Vector instructions per cycle for TC and SEAM-V, together with the SEAM-V-to-TC ratio. (b) Average scalar and vector instructions per EP and the non-singleton EP ratio.}
    \label{fig:instr_supply_ep_packing}
\end{figure*}

\begin{figure}[t]
    \centering
    \includegraphics[width=0.90\linewidth]{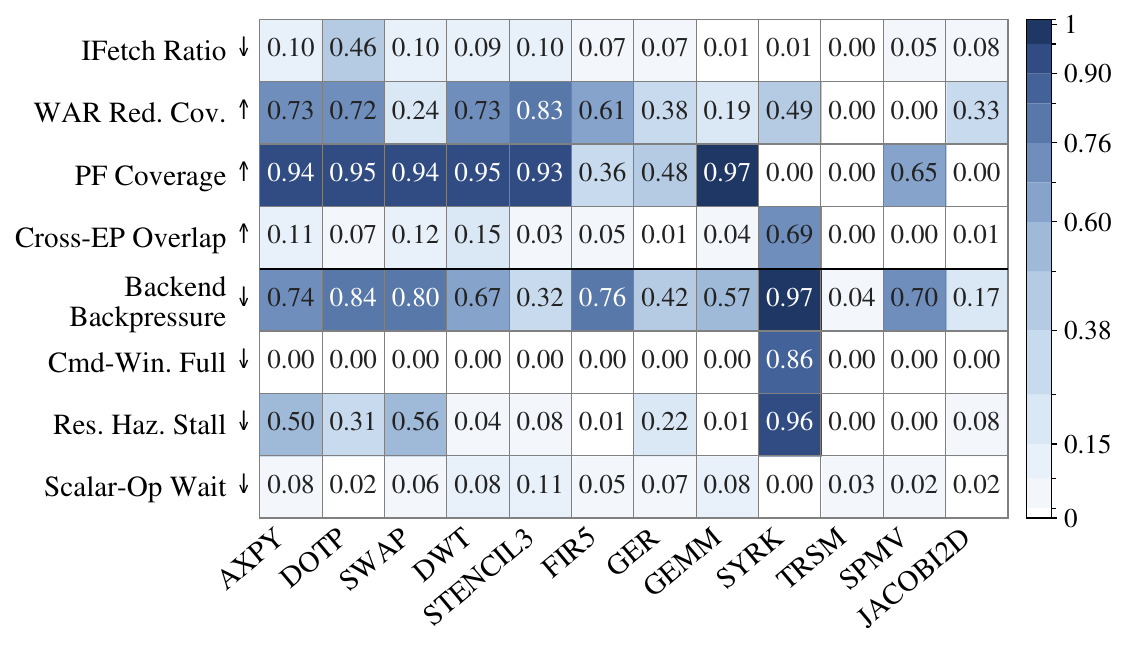}
    \caption{Mechanism activity and residual execution pressure in the full SEAM-V configuration. Arrows indicate lower overhead or pressure (\(\downarrow\)) and greater mechanism activity (\(\uparrow\)).}
    \label{fig:backend_semantic_pressure}
\end{figure}

This section uses microarchitectural counters to characterize the hardware activity associated with the performance results of the full SEAM-V configuration. Whereas Section~\ref{sec:eval_ablation} quantifies the incremental contributions of the key mechanisms through configuration differences, this section examines their actual behavior along the complete execution path. We focus on the vector instruction throughput enabled by local instruction supply, the organization of scalar and vector instructions within EPs, and the activity of request-bound prefetching, source-lifetime-aware scheduling, and cross-EP overlap. Fig.~\ref{fig:instr_supply_ep_packing} analyzes instruction throughput and EP organization, while Fig.~\ref{fig:backend_semantic_pressure} summarizes mechanism activity and residual execution pressure.

Fig.~\ref{fig:instr_supply_ep_packing}(a) shows that the SEAM-V vector instruction rate exceeds the corresponding TC rate for every representative kernel, demonstrating that SEAM-V delivers vector instructions to the backend more continuously. The throughput ratios of DWT, STENCIL3, AXPY, and SWAP reach approximately \(1.92\times\), \(1.75\times\), \(1.64\times\), and \(1.60\times\), respectively, while FIR5 reaches approximately \(1.46\times\). DOTP, GER, and TRSM show comparatively limited improvements. Fig.~\ref{fig:instr_supply_ep_packing}(b) further shows the EP organization characteristics of the evaluated kernels. FIR5 has the largest average EP width and a non-singleton EP ratio close to 1.00, consistent with the local grouping opportunities provided by its multi-load filtering structure. DWT, AXPY, DOTP, and SWAP also exhibit substantial multi-instruction EP activity. STENCIL3 has a comparatively lower non-singleton EP ratio, while SYRK, TRSM, SPMV, and JACOBI2D have smaller average EP widths, indicating fewer opportunities for local instruction aggregation in these kernels.

Fig.~\ref{fig:backend_semantic_pressure} shows the activity of the key mechanisms in the full SEAM-V configuration and how this activity varies across kernels. The upper rows report the instruction fetch ratio, WAR reduction coverage, prefetch coverage, and cross-EP overlap. The instruction fetch ratio is defined as the task-level instruction fetch traffic of SEAM-V relative to that of TC. All evaluated kernels remain below approximately 0.47, reflecting the reduction in repeated instruction fetching provided by local instruction supply and buffer-resident replay. WAR reduction coverage represents the fraction of sequencer commands for which WAR constraints are reduced through source-lifetime release. STENCIL3 has the highest coverage at approximately 0.83. AXPY, DOTP, and DWT each reach approximately 0.72, while FIR5 reaches approximately 0.61, indicating that these kernels make substantial use of source-read completion to release pure WAR dependences early. AXPY, DOTP, SWAP, DWT, STENCIL3, and GEMM all achieve prefetch coverage above 0.92, while GER, FIR5, and SPMV show partial coverage, indicating that the evaluated kernels exploit request-bound prefetching to different degrees. Cross-EP overlap remains relatively low for most kernels. DWT, SWAP, and AXPY reach approximately 0.15, 0.12, and 0.11, respectively, whereas SYRK reaches approximately 0.69, showing that buffered vector early issue creates substantial cross-EP overlap for SYRK but only limited overlap for most other kernels. These activity trends are consistent with the incremental contributions of \texttt{PF} and \texttt{Haz} observed in Section~\ref{sec:eval_ablation}, although mechanism activity does not directly measure final performance contribution. For example, DOTP has high prefetch coverage but obtains almost no standalone benefit from \texttt{PF}, while SYRK exhibits high cross-EP overlap but also experiences substantial backend pressure.

The pressure indicators further reveal the remaining execution constraints in the complete design. SYRK has backend backpressure and command window full ratios of approximately 0.97 and 0.86, respectively, indicating that requests cannot be drained from the command window quickly enough when backend acceptance is constrained, which in turn propagates buffering pressure upstream. DOTP, SWAP, FIR5, AXPY, and SPMV also exhibit noticeable backend backpressure. Residual hazard stall excludes RAW stalls and reflects the remaining WAR/WAW pressure after pure WAR release. Its values are highest for SYRK, SWAP, and AXPY at approximately 0.96, 0.56, and 0.50, respectively, followed by DOTP and GER. FIR5 combines rich EP organization opportunities with substantial backend backpressure, showing that improved front-end instruction organization does not necessarily eliminate backend acceptance limits. Scalar-operand wait remains relatively low across the evaluated kernels, with all values at or below approximately 0.11. Its overall effect on execution progress is therefore limited, and the principal remaining pressures arise from backend backpressure and register dependence stalls.

Overall, Figs.~\ref{fig:instr_supply_ep_packing} and~\ref{fig:backend_semantic_pressure} capture the microarchitectural behavior of SEAM-V, from local instruction supply and EP formation to execution in the vector backend. Local instruction supply improves vector instruction organization and delivery, backend-visible packet semantics support request-bound prefetching, source-lifetime-aware scheduling uses actual source-read completion to shorten pure WAR dependences, and buffered vector early issue provides controlled cross-EP overlap when the required safety conditions are satisfied. EP organization opportunities and mechanism activity differ substantially across kernels. After instruction supply is improved, backend backpressure, command window capacity, and residual dependence pressure remain the primary constraints, while scalar operand waits have a more limited effect.

\subsection{Hardware Cost and Physical Design}
\label{sec:eval_hardware_cost}

We synthesize TC and SEAM-V using Synopsys Design Compiler, targeting TSMC 28-nm HPC+~\cite{tsmc28hpcplus} at TT/0.9\,V/25\(^{\circ}\)C under a 1-GHz synthesis target. Power is estimated using Synopsys Power Compiler with SAIF activity generated from gate-level simulations~\cite{powercompiler2022}. As shown in Table~\ref{tab:synthesis_area}, the standard cell logic area of SEAM-V increases by 16.76\%, with combinational and noncombinational areas increasing by 19.71\% and 10.38\%, respectively. The larger combinational increase is consistent with the additional control and selection logic required by the execution and coordination mechanisms. Nevertheless, the reported total cell area increases by only 4.29\%, indicating a limited overall area increase after integrating the complete hybrid-decoupled execution path and the two backend coordination mechanisms.

\begin{table}[t]
    \centering
    \caption{Post-synthesis area comparison between TC and SEAM-V.}
    \label{tab:synthesis_area}
    \scriptsize
    \setlength{\tabcolsep}{4.0pt}
    \renewcommand{\arraystretch}{1.08}

    \begin{tabular}{@{}lrrr@{}}
        \toprule
        Area metric (\(\mathrm{mm}^{2}\))
            & TC
            & SEAM-V
            & \(\Delta\) (\%) \\
        \midrule
        Combinational
            & 0.7744
            & 0.9271
            & +19.71 \\
        Noncombinational
            & 0.3583
            & 0.3955
            & +10.38 \\
        Standard cell logic
            & 1.1327
            & 1.3226
            & +16.76 \\
        \midrule
        Macro/black box
            & 1.7953
            & 1.7310
            & -3.58 \\
        Total cell area
            & 2.9280
            & 3.0536
            & +4.29 \\
        \bottomrule
    \end{tabular}
\end{table}

\begin{figure}[t]
    \centering
    \includegraphics[width=0.55\linewidth]{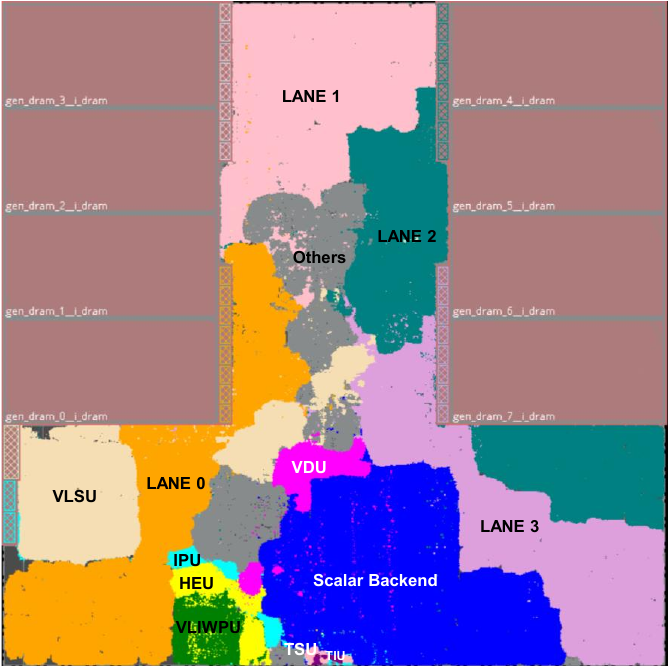}
    \caption{Placed layout of SEAM-V, showing the physical organization of the hybrid-decoupled execution path, local scalar backend, and vector backend.}
    \label{fig:seamv_placed_layout}
\end{figure}

Table~\ref{tab:power_energy} shows that the geometric mean normalized power of SEAM-V increases by 17.30\%, whereas task energy decreases by 12.70\%; 11 of the 12 evaluated kernels achieve lower task energy. This result indicates that shorter execution time offsets the additional hardware activity in most workloads. JACOBI2D exhibits the largest power increase but still reduces task energy by 16.00\%, whereas DOTP shows a 5.60\% energy regression because of its limited performance gain. Energy efficiency therefore depends on whether the added mechanisms provide sufficient cycle reduction. Fig.~\ref{fig:seamv_placed_layout} presents the physical organization of the hybrid-decoupled execution path, local scalar backend, and Ara-based vector backend in the placed design. The resulting module distribution reflects the architectural separation between local instruction and control processing and the vector execution backend. Overall, SEAM-V trades limited area and power overheads for more continuous instruction supply, greater memory--computation overlap, and fewer conservative dependence waits, thereby shortening task execution time and reducing task-level energy for most workloads.

\begin{table}[t]
    \centering
    \caption{SAIF-based post-synthesis power and energy comparison.}
    \label{tab:power_energy}
    \scriptsize
    \setlength{\tabcolsep}{2.4pt}
    \renewcommand{\arraystretch}{1.05}

    \begin{tabular}{@{}lrr@{\hspace{6pt}}rr@{}}
        \toprule
        & \multicolumn{2}{c}{Total power (mW)}
        & \multicolumn{2}{c}{Normalized to TC} \\
        \cmidrule(lr){2-3}
        \cmidrule(l){4-5}
        Kernel
            & TC
            & SEAM-V
            & \(P/P_{\mathrm{TC}}\)
            & \(E/E_{\mathrm{TC}}\) \\
        \midrule
        AXPY     & 172.51 & 197.95 & 1.148 & 0.837 \\
        DOTP     & 157.80 & 178.46 & 1.131 & 1.056 \\
        SWAP     & 164.29 & 184.37 & 1.122 & 0.807 \\
        DWT      & 193.50 & 232.68 & 1.203 & 0.779 \\
        STENCIL3 & 171.03 & 206.93 & 1.210 & 0.803 \\
        FIR5     & 177.40 & 209.78 & 1.183 & 0.873 \\
        GER      & 198.06 & 209.36 & 1.057 & 0.858 \\
        GEMM     & 213.20 & 264.39 & 1.240 & 0.815 \\
        TRSM     & 178.91 & 194.54 & 1.087 & 0.996 \\
        SYRK     & 147.13 & 171.21 & 1.164 & 0.963 \\
        JACOBI2D & 171.29 & 239.33 & 1.397 & 0.840 \\
        SPMV     & 150.03 & 175.85 & 1.172 & 0.892 \\
        \midrule
        GeoMean  & -- & -- & 1.173 & 0.873 \\
        \bottomrule
    \end{tabular}

    \vspace{1mm}
    \begin{minipage}{0.98\columnwidth}
        \scriptsize
        Task energy equals average power multiplied by the SAIF duration.
    \end{minipage}
\end{table}

\section{Conclusion}
\label{sec:conclusion}

This paper presents SEAM-V, a hybrid-decoupled RVV processor that uses task-level decoupling, local instruction supply, and HINT-guided EP formation to organize a continuous scalar/vector instruction stream. Request-bound prefetching and source-lifetime-aware scheduling improve memory access coordination and the early release of pure WAR dependences, respectively. Software provides program structure and safety constraints, while vector dependences and request progression remain dynamically managed by the backend, thereby preserving standard RVV semantics. Cycle-accurate RTL evaluation shows that SEAM-V achieves a geometric mean speedup of \(1.38\times\) across 17 representative kernel configurations; across seven ablation kernels, the full configuration reduces geometric mean normalized cycles to 0.64. Synthesis and power analysis show a 4.29\% increase in total cell area and a 12.70\% reduction in task energy. The results further indicate that, after instruction supply is improved, backend backpressure, buffer capacity, and residual dependence pressure may become the dominant constraints. Overall, the collaboration in which software provides program structure and memory intent while hardware retains runtime decision-making offers a viable path toward cross-layer optimization in scalable RVV processors.

\bibliographystyle{IEEEtran}
\bibliography{refs}

\end{document}